\documentclass[aps,prl,showpacs,twocolumn,showkeys,amssymb,amsmath,superscriptaddress,reprint]{revtex4-1}

\usepackage{amsmath}
\usepackage{graphicx}
\usepackage{color}

\usepackage{verbatim}

\begin{document}

\title{Susceptibilities for the M\"{u}ller-Hartmann-Zitartz countable infinity of phase transitions on a Cayley tree}

\author{Auditya Sharma}
\affiliation{School of Chemistry, The Faculty of Exact Sciences, Tel Aviv University, Tel Aviv 69978, Israel}

\begin{abstract}
We obtain explicit susceptibilities for the countable infinity of phase transition temperatures of M\"{u}ller-Hartmann-Zitartz on a Cayley tree.
The susceptibilities are a product of the zeroth spin with the sum of an appropriate set of averages of spins on the outermost layer of the tree.
A clear physical understanding for these strange phase transitions emerges naturally. In the thermodynamic limit, the susceptibilities tend to zero above the 
transition and to infinity below it.
\end{abstract}

\maketitle

\section{Introduction}
Forty years ago, M\"{u}ller-Hartmann and
Zitartz~\cite{MHZ:74,MHZ:75,MH:77} showed that the Ising ferromagnet
on a Cayley tree has a countable infinity of phase transitions of an
unusual type. Specifically, in the low temperature ordered phase, when one traverses
vertically across the zero-field region in the magnetic field - temperature phase diagram, they 
showed that the \emph{order} of the phase transition can be \emph{anything} between $1$ and $\infty$ 
depending on the value of the temperature at that point. Therefore, they dubbed them phase transitions
of continuous order. By identifying the values of temperature at which the order acquires integer values,
they also extracted a countable infinity of phase-transitions in the low-temperature phase. The MHZ work constitutes 
one of the classic examples of a strange phase transition. Most of their work is rather technical and 
mathematical though, and a transparent physical understanding of the meaning of these strange
phase transitions would be useful. 

Although criticized as unphysical, the Cayley tree (or the close cousin called Bethe lattice) is a popular geometric
structure on which innumerable studies continue to be carried out in current times~\cite{ostilli2012cayley,accardi2013quantum,rozikov2013gibbs,mukhamedov2013dynamical,khorrami2014autonomous,ganikhodjaev2013blume,changlani2013emergent}. One reason is of course the availability of exact solutions for some models. Another important, not often emphasized reason is that it simultaneously contains one-dimension like properties and infinite-dimension like properties, thus providing an interesting framework for theoretical explorations. For example, the partition function
of the ferromagnet on a Cayley tree~\cite{baxter2014solving,eggarter1974cayley} is identical to that of the one-dimensional chain (barring an inconsequential constant factor), and yet
a phase transition exists in the Cayley showing infinite-dimensional character. 

MHZ extract the order of the phase transition at various points in the low-temperature phase by studying the leading-order singularities of the full 
free-energy as a function of field as it is taken to the zero limit. Here we show that a direct transparent understanding of these phase transitions
maybe obtained by the consideration of the pure zero-field model, building on a `memory-approach' that we emphasized in recent work~\cite{magansharma:2013}. 
We were also partially propelled by the recent exact solution of a one-dimensional long-range ferromagnet which admits an unusual phase transition of 
a different type, namely mixed-order, which can simultaneously show a discountinuous jump in maganetization (first-order like), \emph{and} a diverging
correlation-length (second-order like)~\cite{barmukamelprl:2014,barmukameljstat:2014}. 

\section{The Countable infinity of phase transitions of MHZ}
\begin{figure}
\includegraphics[width=0.8\hsize, trim = 0 140mm 0 0, clip = true]{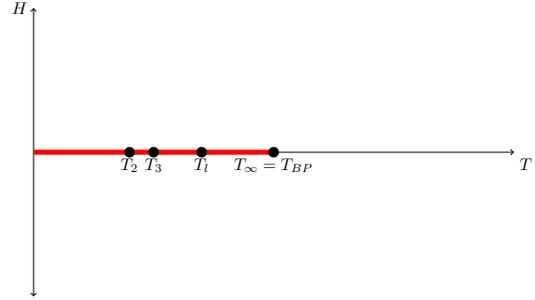}
\caption{Schematic of the countable infinity of phase transitions of MHZ. The dots correspond to the countable infinity of 
phase transitions (as one traverses vertically by varying $H$ through $0$ at a given temperature in the ordered phase) of integer order.}
\label{fig1}
\end{figure}
Fig~\ref{fig1} carries a schematic of the phase diagram of MHZ. The gist of their approach is to make a careful detailed study 
of the free energy $F(H,T)$. In a generic ferromagnetic system with a phase transition, $F(H,T)$ is analytic at all points except in the region shown in red: $[H=0, 0\le T\le T_{BP}]$.
In this limit the free energy is given by~\cite{MHZ:75} 
\begin{equation}
F(H,T) = F(0,T) + f_{reg}(H^{2},T) + A(T) |H|^{\kappa}, H \to 0, 
\end{equation}
where the regular part is a function of $H^{2}$ because of symmetry, and the leading singular part $A(T)$ shows a power law
behavior. The unusual aspect of the Cayley tree lies in the fact that the critical exponent $\kappa$ varies continuously 
from $1$ at $T=0$ to $\infty$~\cite{MHZ:75} at the usual phase transition called the Bethe-Peierls transition $T_{BP}$. This behaviour is in sharp 
contrast to most commonly encountered phase transitions where $\kappa$ remains a constant ($=1$ for a first order transition, $2$ for 
a second order transition and so on). By studying the points at which $\kappa$ takes integer values, they identify a countable infinity 
of phase transitions which fit into the Ehrenfest classification of phase transitions of integer power-order.

Here we point out that in fact a simple (albeit unusual) set of susceptibilities identify the countable infinity of phase
transitions, and indeed a clear physical picture of the meaning behind
the phase transitions comes naturally out of them. The susceptibilities turn out to be a product of the zeroth spin with the sum of averages
of appropriately grouped spins in the outermost layer of the tree. Furthermore, we can work entirely with the zero-field model, with 
no requirement of complicated procedures or mathematical methods related to the application of a tiny field followed by taking the zero-field limit. 

\section{The Model}
\begin{figure}
\includegraphics[width=\hsize, trim = 0 145mm 0 0, clip = true]{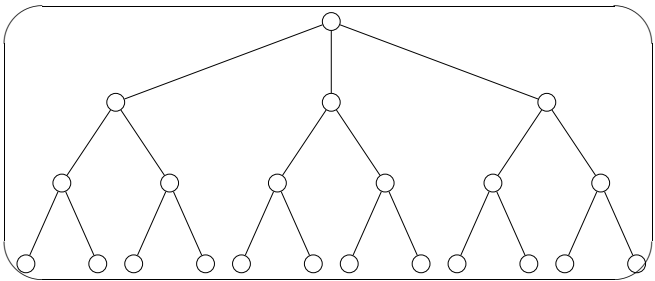}
\caption{Figure shows a Cayley tree of depth $n=3$ and coordination number $z=3$.}
\label{fig2}
\end{figure}
 Following the notation of a recent piece of work involving the author~\cite{magansharma:2013}, we consider the
following Hamiltonian:
\begin{equation}
\mathcal{H} = -J\sum_{\langle i, j \rangle} \sigma_{i} \sigma_{j} + H\sum_{j}\sigma_{j},
\label{Ham}
\end{equation}
where the sum involves pairs of spins which are adjacent on the tree
(Fig.~\ref{fig2}) with coordination number $z$ and depth
$n$. $\sigma_{i}$ are Ising variables which can take values $\pm 1$,
and J is taken to be positive to make it a ferromagnet. The solution, when the external field
is $H = 0$, is trivial and we quickly recall it. We introduce new bond variables $\theta_{ij} =
\sigma_{i}\sigma_{j}$. The $\theta_{ij}$ can take values $\pm 1$,
which make them effectively spin variables too. Specifying all the
$\theta_{ij}$, and the spin at the root of the tree $\sigma_{0}$,
completely defines the system. The Hamiltonian then takes the
following simple form
\begin{equation}
\mathcal{H} = -J\sum_{\langle i, j \rangle} \theta_{ij}.
\end{equation}
With the problem now rehashed into one with non-interacting spins
under the influence of an external magnetic field, the partition
function is readily written down~\cite{eggarter1974cayley}:
\begin{equation}
\mathcal{Z}(J)_{Cayley} = 2(2\cosh(\beta J))^{N_{b}} = 2(2\cosh(\beta J))^{N-1},
\end{equation}
where $N_{b}$ is the number of bonds and $N$ is the number of spins,
and $\beta= 1/T$ is the inverse temperature as usual in equilibrium
statistical mechanics.  It follows
directly~\cite{mukamel:74,matsuda1974infinite,morita1975susceptibility,von1974phase,magansharma:2013} that the correlation
function between any two spins $\sigma_{i}, \sigma_{j}$ is given by:
\begin{equation}
\langle \sigma_{i}\sigma_{j}\rangle = \tanh(\beta J)^{d_{ij}} = a^{d_{ij}},
\end{equation}
where we define $a \equiv \tanh(\beta J)$ for convenience, and
$d_{ij}$ is the distance of the (unique) shortest path between the
points $i,j$.

\begin{figure}
\includegraphics[width=\hsize]{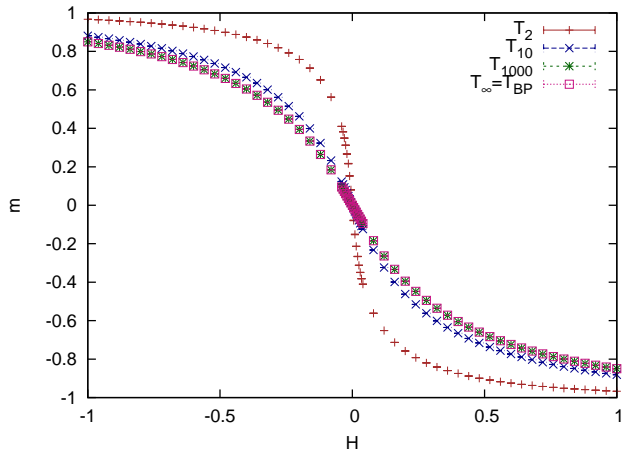}
\caption{A plot of the magnetization $m = \frac{1}{N}\sum_{i}\langle\sigma_{i}\rangle$ versus magnetic field from Monte Carlo simuations of a finite-sized sytem
of coordination number $z=3$ and depth $n=8$. Data are shown for different runs at the MHZ transition temperatures $T_{2}, T_{10}, T_{1000}, T_{\infty}\equiv T_{BP}$.
At higher transition temperatures, the curve becomes smoother and smoother indicating that in the thermodynamic limit the singularity would occur at a higher order differentiation
with respect to $H$. Since our system is finite, the data are practically indistinguishable for $T_{1000}$ and $T_{\infty}$.
}
\label{fig3}
\end{figure}

When the external field $H$ is non-zero, there is no simple closed-form expression, however MHZ~\cite{MHZ:74} write down an
infinite-series expansion for the free energy $F(H,T)$, and by a careful, elaborate study of the order in field at which the leading
singularity occurs as one crosses zero-field at low-temperature, they obtain the following countable infinity of transition temperatures:
\begin{equation}
\label{eqn:Tl}
T_{l}  = \frac{J}{\tanh^{-1}(\frac{1}{\gamma^{(l-1)/l}})},
\end{equation}
$l=1,2,\cdots,\infty$, with $l$ being identified as the order of the phase transition within the Ehrenfest scheme. $T_{1} = 0$ has a first order phase transition
and $T_{\infty} = T_{BP}$ with order infinity is the so-called `Bethe-Peierls' phase transition which is the temperature at which the system first orders as it is cooled down
from high temperature. Fig~\ref{fig3} shows a Monte Carlo simulation carried out on a finite-sized system of depth $n=8$ and coordination number $z=3$. We
studied the dependence of magnetization as one crosses from negative to positive magnetic field at some of the transition temperatures of MHZ. It seems plausible
that in the thermodynamic limit, at a higher transition temperature, a greater derivative of the magnetization with respect to field would diverge at $H=0$. Although the
simulations were run on a finite-system the data at $T_{2}$ display a considerably sharp drop near $H=0$ indicative of the diverging derivative, since it is a second order
phase transition.

Here we show that the above countable infinity of transition temperatures may be directly obtained from the zero-field model bypassing the
elaborate complicated methods of studying the infinite series and the order of divergence of the free energy in the limit of $H\to 0$. We do this by an explicit
construction of a special set of `susceptibilities' for the transitions $T_{l}$.
\section{Construction of the susceptibilities for the MHZ transition temperatures}
In order to construct the susceptibilities for the MHZ transition
temperatures, we choose the depth of the lattice to be of the form $n
= lp$, where $l$ can take values $1,2,3,\cdots,\infty$. Since we
are interested in the $p\to\infty$ limit, no loss of generality is
incurred. The number of spins in the $n^{th}$ layer of the tree is
$N_{n} = z(z-1)^{n-1} \equiv (1+\frac{1}{\gamma})\gamma^{n}$, where we
have defined $\gamma \equiv (z-1)$ for convenience. Let us denote the
$n^{th}$ layer spins by
$\sigma_{n,1},\sigma_{n,2},\cdots,\sigma_{n,N_{n}}$ in order from left
to right, as can be visualized in Fig.~\ref{fig2}. Next, we group the
first $M_{p}\equiv\gamma^{p}$ spins of the $n^{th}$ layer and call
their average $\tilde{\sigma}_{n,1} =
\frac{1}{M_{p}}\sum_{i=1}^{M_{p}}\sigma_{n,i}$; we then group the
second $M_{p}$ spins of the $n^{th}$ layer and call their average
$\tilde{\sigma}_{n,2} = \frac{1}{M_{p}}\sum_{i=M_{p}+1}^{2M_{p}}\sigma_{n,i}$, and so on. By
this procedure we form $K_{n,p} \equiv \frac{N_{n}}{M_{p}}$ spin averages:
\begin{align}
\tilde{\sigma}_{n,j} &= \frac{1}{M_{p}}\sum_{i=(j-1)M_{p}+1}^{jM_{p}}\sigma_{n,i}\\
j &= 1,2,3,\cdots,K_{n,p}.\nonumber
\end{align}

Now we are ready to write down the susceptibilities. Recalling that $\sigma_{0}$ is the spin at the root node, the
susceptibilities are simply given by:
\begin{align}
\mathcal{X}_{l} = \sigma_{0}\sum_{j=1}^{K_{n,p}}\tilde{\sigma}_{n,j},
\end{align}
which is our main result. To see that this leads to the MHZ transition temperatures, let us invoke the two-point 
correlation functions from the last section to compute the expectation value:
\begin{align}
\langle \mathcal{X}_{l}\rangle &= \langle \sigma_{0}\sum_{j=1}^{K_{n,p}}\tilde{\sigma}_{n,j} \rangle\\
                     &= \sum_{j=1}^{K_{n,p}}\frac{1}{M_{p}}\sum_{i=(j-1)M_{p}+1}^{jM_{p}}\langle\sigma_{0}\sigma_{n,i}\rangle\nonumber\\
                     &= \sum_{j=1}^{K_{n,p}}\frac{1}{M_{p}}\sum_{i=(j-1)M_{p}+1}^{jM_{p}} a^{n}\nonumber\\
                     &= {K_{n,p}}a^{n}\nonumber\\
                     &= (1+\frac{1}{\gamma})[\gamma^{(l-1)}a^{l}]^{p}\nonumber.\\
\end{align} 
Therefore we see that as $p\to\infty$, 
\begin{equation}
\langle \mathcal{X}_{l}\rangle \to \begin{cases}
  0 & \text{if $\gamma^{(l-1)}a^{l} < 1$} \\ 
  \infty & \text{if $\gamma^{(l-1)}a^{l} > 1$}.
\end{cases}
\end{equation}
$\gamma a^{\frac{l}{(l-1)}} = 1$ thus defines the $l^{th}$
transition temperature. These are precisely the transition
temperatures of MHZ as given in Eqn.~\ref{eqn:Tl}~\cite{MHZ:74,falk1975ising}.  It is worth pointing out that for
$l=2$, this yields the so-called (because of the appearance of the
same in the disordered version of the same
Hamiltonian~\cite{magansharma:2013}) `spin-glass' transition
temperature $T_{SG} = \frac{J}{\tanh^{-1}(\frac{1}{\sqrt{\gamma}})}$,
and the $l\to\infty$ limit yields the Bethe-Peierls transition
temperature $T_{BP}=\frac{J}{\tanh^{-1}(\frac{1}{\gamma})}$.

\section{Conclusions}
We have introduced a set of `susceptibilities' that help identify the mysterious MHZ
transition temperatures on the Cayley tree in a transparent manner. They are given by the product of
the zeroth spin with an appropriately averaged sum of spins from the
outermost layer in a Cayley tree. A clear physical understanding of the phase-transitions emerges
naturally. We observe that our susceptibilities have the feature that in the thermodynamic limit, they are zero above the phase
transition, but tend to infinity below it. We are able also to 
identify the second-order phase transition $T_{2}$ as the phase transition known as the spin-glass 
phase transition in the literature.

Furthermore, although we have concentrated on the
Ising ferromagnet here, the susceptibilites defined here are
primarily attached to the geometry of the lattice. Therefore, they should be applicable much
more generally: for example with $m$-component vector spins and the
bonds could be ferromagnetic or antiferromagnetic or disordered. Quantum
models should display similar transitions as well: a detailed investigation
of various models from this perspective would be desirable.

Finally, we remark that the statistical mechanics problem on the tree has been connected with the problem of 
reconstruction of information on trees in formal, extensive studies~\cite{mezard2006reconstruction,mossel2004survey}. It would
be interesting to understand if and how the MHZ countable infinity of phase transitions would fit into this generalized problem,
which might be of interest to a broader community.
\begin{acknowledgments}
We gratefully acknowledge numerous discussions and collaborative
work with Javier Martinez Mag\'an, which paved the way towards the
formulation of this project. We are indebted to David Mukamel for enlightening discussions that helped crystallize
the story. This work was supported by The Center for Nanoscience and Nanotechnology at Tel Aviv University and the PBC Indo-Israeli Fellowship.
We are grateful to the anonymous referees for certain references and for constructive suggestions.
\end{acknowledgments}

\bibliography{MHZ_susc}

\end{document}